\documentclass[a4paper,conference]{IEEEtran}

\usepackage{ifthen}
\newboolean{arxiv}
\setboolean{arxiv}{true}

\usepackage{amsmath,amssymb}
\usepackage{amsthm} 
\usepackage{ascmac}
\usepackage{bbm}
\usepackage{bm} 
\usepackage{braket}
\usepackage{cite}
\usepackage{dcolumn} 
\usepackage{enumerate}
\usepackage{enumitem}
\usepackage{framed}
\usepackage{graphicx} 
\ifthenelse{\boolean{arxiv}}{}{\usepackage[dvipdfmx]{hyperref}}
\usepackage{ifthen}
\usepackage{longtable}
\usepackage{mathtools}
\usepackage{multirow}
\usepackage{overpic}
\usepackage{pict2e}
\ifthenelse{\boolean{arxiv}}{}{\usepackage{pxjahyper}}
\usepackage[usestackEOL]{stackengine}
\usepackage{txfonts} 
\usepackage{xparse}

\usepackage{color}
\usepackage[most]{tcolorbox}

\newcommand{\hypercolor}{blue}
\hypersetup{
  colorlinks,
  citecolor=\hypercolor,
  linkcolor=\hypercolor,
  urlcolor=\hypercolor,
  bookmarksopen=true,
  bookmarksopenlevel=4,
  bookmarksnumbered
}
\ifthenelse{\boolean{arxiv}}{}{\mathtoolsset{showonlyrefs,showmanualtags}}

\usepackage{bm}

\theoremstyle{definition}

\tcbuselibrary{theorems}
\tcbset{
  code={},
  colback=white,
  noparskip,
  before=\par\bigskip\noindent,
  coltitle=black,fonttitle=\upshape\bfseries,
  breakable,frame empty,interior empty,colframe=white,
  description delimiters parenthesis,
  enhanced,theorem style=plain,
  boxsep=0mm,
  leftrule=0mm,rightrule=0mm,
  left=0mm,right=0mm,top=0mm,bottom=0mm,
  before skip=10pt,after skip=10pt,
  posstyle/.style={},
  thmstyle/.style={},
}
\newtcbtheorem{postulate}{Postulate}{posstyle}{pos}
\newtcbtheorem{thm}{Theorem}{thmstyle}{thm}
\newtcbtheorem[use counter from=thm]{lemma}{Lemma}{thmstyle}{lemma}
\newtcbtheorem[use counter from=thm]{proposition}{Proposition}{thmstyle}{pro}
\newtcbtheorem[use counter from=thm]{cor}{Corollary}{thmstyle}{cor}

\newcounter{ex}

\ExplSyntaxOn
\NewDocumentEnvironment{ex}{o}
 {
  \refstepcounter{ex}
  \par\smallskip
  \noindent
  \textbf{\IfNoValueTF{#1}{Example}{Example~of~#1}~}
 }
 {\par\smallskip
 }
\ExplSyntaxOff

\newcommand{\ident}{\hat{1}}

\newcommand{\Real}{\mathbb{R}}

\newcommand{\T}{\mathsf{T}}
\DeclareMathOperator{\Tr}{Tr}

\DeclareMathAlphabet{\mymathbb}{U}{BOONDOX-ds}{m}{n}
\newcommand{\zero}{\mymathbb{0}}

\newboolean{figure}
\setboolean{figure}{true}
\newcommand{\InsertPDF}[1]{\iffigure\includegraphics[scale=1.0]{#1}\fi}
\ifthenelse{\boolean{figure}}{}{}

\renewcommand{\ident}{\mathbbm{1}}
\renewcommand{\c}{\circ}
\newcommand{\ot}{\otimes}
\newcommand{\cross}{\times}
\newcommand{\pp}[1]{\frac{\partial}{\partial #1}}
\newcommand{\ppp}[2]{\frac{\partial^2}{\partial #1 \partial #2}}

\let\protect\relax
{\catcode`\|=\active
  \xdef\Craket{\protect\expandafter\noexpand\csname Craket \endcsname}
  \expandafter\gdef\csname Craket \endcsname#1{\begingroup
     \ifx\SavedDoubleVert\relax
       \let\SavedDoubleVert\|\let\|\BraDoubleVert
     \fi
     \mathcode`\|32768\let|\BraVert
     \left({#1}\right)\endgroup}
}

\definecolor{gray}{RGB}{128,128,128}

\newcommand{\Discard}[1]{}

\begin{document}
\title{Matrix differentiation with diagrammatic notation}
\author{
 \IEEEauthorblockN{Kenji~Nakahira}
 \IEEEauthorblockA{
  Quantum Information Science Research Center, \\
  Quantum ICT Research Institute, Tamagawa University \\
  6-1-1 Tamagawa-gakuen, Machida, Tokyo 194-8610 Japan\\
  {\footnotesize\tt E-mail: nakahira@lab.tamagawa.ac.jp} 
  \vspace*{-2.64ex}}
}

\maketitle

\begin{abstract}
 We propose a diagrammatic notation for matrix differentiation.
 Our new notation enables us to derive formulas for matrix differentiation
 more easily than the usual matrix (or index) notation.
 We demonstrate the effectiveness of our notation through several examples.
\end{abstract}

\IEEEpeerreviewmaketitle

\section{Introduction} \label{sec:intro}

Matrix differentiation (or matrix calculus) is widely accepted as an essential tool in various fields
including estimation theory, signal processing, and machine learning.
Matrix differentiation provides a convenient way to collect
the derivative of each component of the dependent variable
with respect to each component of the independent variable,
where the dependent and independent variables can be a scalar, a vector, or a matrix.
However, the usual matrix (or index) notation often suffers from cumbersome calculations
and difficulty in the intuitive interpretation of the final results.
It is known that diagrammatic representations using string diagrams can be successfully applied
in linear algebra (see \cite{Coe-2010} and references therein).
In this paper, we provide a simple diagrammatic approach
to derive useful formulas for matrix differentiation.

Here we mention some related work.
In Ref.~\cite{Kim-Oh-Kim-2019}, the way of graphically representing the del operator (i.e., $\nabla$)
is presented, in which calculations are limited to the case of three-dimensional Euclidean space.
Reference~\cite{Tou-Yeu-Fel-2021} presents a diagrammatic notation for manipulating tensor
derivatives with respect to one parameter.
We adopt a similar notation to those given in these references.

\section{Definition of matrix differentiation}

Let $\Real$ be the set of all real numbers
and $\Real^{m \times n}$ be the set of all $m \times n$ real matrices.
Also, let $\{ \ket{i} \}_{i=1}^m$ denote the standard basis of $\Real^m$.
We are concerned only with finite-dimensional real Hilbert spaces.
Given a map $f$ from $\Real^{m \times n}$ to $\Real$ and
a matrix $X \in \Real^{m \times n}$ of independent variables,
we denote the $m \times n$ real matrix with $(i,j)$-th component
$\pp{X_{i,j}} f(X)$ by $\pp{X} f(X)$,
where $X_{i,j} \coloneqq \braket{i|X|j}$ is the $(i,j)$-th component of $X$.
We have
\begin{alignat}{1}
 \pp{X} f(X) &=
 \sum_{i=1}^m \sum_{j=1}^n \pp{X_{i,j}} \ket{i} \bra{j} f(X) \nonumber \\
 &=
 \begin{bmatrix}
  \pp{X_{1,1}} f(X) & \pp{X_{1,2}} f(X) & \cdots & \pp{X_{1,n}} f(X) \\
  \pp{X_{2,1}} f(X) & \pp{X_{2,2}} f(X) & \cdots & \pp{X_{2,n}} f(X) \\
  \vdots & \vdots & \ddots & \vdots \\
  \pp{X_{m,1}} f(X) & \pp{X_{m,2}} f(X) & \cdots & \pp{X_{m,n}} f(X) \\
 \end{bmatrix}.
 \label{eq:fX}
\end{alignat}
In the special case of $n = 1$, $X$ is a column vector, which is denoted by $\ket{x}$.
In this case, we have
\begin{alignat}{1}
 \pp{\ket{x}} f(\ket{x}) &=
 \sum_{i=1}^m \pp{x_i} \ket{i} f(\ket{x})
 =
 \begin{bmatrix}
  \pp{x_{1}} f(\ket{x}) \\
  \pp{x_{2}} f(\ket{x}) \\
  \vdots \\
  \pp{x_{m}} f(\ket{x}) \\
 \end{bmatrix},
\end{alignat}
where $x_i \coloneqq \braket{i|x}$.

A similar notation is used when $f$ is a map
from $\Real^{m \times n}$ to $\Real^{m' \times n'}$.
For such $f$, $\pp{X} f(X)$ is an $m \times n \times m' \times n'$ fourth-order tensor
with components $\{ \pp{X_{i,j}} \braket{i'|f(X)|j'} \}_{i,j,i',j'}$.
This can be written as the following $mm' \times nn'$ matrix:
\begin{alignat}{1}
 \pp{X} f(X) &=
 \begin{bmatrix}
  \pp{X_{1,1}} f(X) & \pp{X_{1,2}} f(X) & \cdots & \pp{X_{1,n}} f(X) \\
  \pp{X_{2,1}} f(X) & \pp{X_{2,2}} f(X) & \cdots & \pp{X_{2,n}} f(X) \\
  \vdots & \vdots & \ddots & \vdots \\
  \pp{X_{m,1}} f(X) & \pp{X_{m,2}} f(X) & \cdots & \pp{X_{m,n}} f(X) \\
 \end{bmatrix},
\end{alignat}
where, for each $i$ and $j$, $\pp{X_{i,j}} f(X)$ is the $m' \times n'$ matrix
whose $(i',j')$-th component is $\pp{X_{i,j}} \braket{i'|f(X)|j'}$.

\section{Diagrammatic notation}

In diagrammatic terms, a matrix is represented as a box with an input wire at the bottom
and an output wire at the top.
Column vectors, row vectors, and scalars are regarded as special cases of matrices.
For example, $A \in \Real^{m \times n}$,
$\ket{x} \in \Real^m \coloneqq \Real^{m \times 1}$,
$\ket{y} \in \Real^{m*} \coloneqq \Real^{1 \times m}$,
and $p \in \Real$ are diagrammatically depicted as
\begin{alignat}{1}
 \InsertPDF{component.pdf} ~\raisebox{.7em}{.}
 \label{eq:component}
\end{alignat}
The Hilbert space $\Real^m$ is represented by the wire with label $m$,
while the Hilbert space $\Real$ is represented by `no wire'.
For a scalar, the box will be omitted.
Matrix multiplication and tensor products are represented as
the sequential and parallel compositions, respectively.
The identity matrix $\ident \in \Real^{m \times m}$ is depicted as
\begin{alignat}{1}
 \InsertPDF{matrix_ident.pdf} ~\raisebox{.7em}{.}
 \label{eq:matrix_ident}
\end{alignat}
We often use a special column vector $\ket{\cup_n} \in \Real^n \ot \Real^n$, called a cup,
and a special row vector $\bra{\cap_n} \in \Real^{n*} \ot \Real^{n*}$, called a cap.
The cup $\ket{\cup_n}$ is depicted as
\begin{alignat}{1}
 \InsertPDF{matrix_cup.pdf} ~\raisebox{.1em}{.}
 \label{eq:matrix_cup}
\end{alignat}
The cap $\bra{\cap_n}$ is the transpose of $\ket{\cup_n}$,
which is depicted as
\begin{alignat}{1}
 \InsertPDF{matrix_cap.pdf} ~\raisebox{.1em}{.}
 \label{eq:matrix_cap}
\end{alignat}
We have that, for any $X \in \Real^{m \times n}$,
\begin{alignat}{1}
 \InsertPDF{matrix_cup_cap_X.pdf} ~\raisebox{.1em}{.}
 \label{eq:matrix_cup_cap_X}
\end{alignat}
Indeed, the left equality is obtained from
\begin{alignat}{1}
 \InsertPDF{matrix_cup_cap_X_proof.pdf} ~\raisebox{.1em}{,}
 \label{eq:matrix_cup_cap_X_proof}
\end{alignat}
and the same argument works for the right equality.
Equation~\eqref{eq:matrix_cup_cap_X} implies that
the transpose acts diagrammatically by rotating boxes $180^\circ$.
Substituting $X = \ident$ with Eq.~\eqref{eq:matrix_cup_cap_X} yields
\begin{alignat}{1}
 \InsertPDF{matrix_cup_cap_id.pdf} ~\raisebox{.1em}{.}
 \label{eq:matrix_cup_cap_id}
\end{alignat}
The trace of $X \in \Real^{m \times m}$ satisfies
$\Tr X = \braket{\cap|X \ot \ident|\cup}$, i.e.,
\begin{alignat}{1}
 \InsertPDF{matrix_trace.pdf} ~\raisebox{.1em}{.}
 \label{eq:matrix_trace}
\end{alignat}
We also use the swap matrix $\cross_{n,m}$, depicted by
\begin{alignat}{1}
 \InsertPDF{matrix_cross.pdf} ~\raisebox{.1em}{,}
 \label{eq:matrix_cross}
\end{alignat}
and the matrix called ``spider'', depicted by
\begin{alignat}{1}
 \InsertPDF{matrix_spider.pdf} ~\raisebox{.1em}{.}
 \label{eq:matrix_spider}
\end{alignat}
For details regarding the properties of these matrices, see, e.g., Ref.~\cite{Coe-2010}.

We write $\pp{X} f(X)$ with a map $f:\Real^{m \times n} \to \Real^{m' \times n'}$ as
\begin{alignat}{1}
 \InsertPDF{matrix_dif_fX.pdf} ~\raisebox{.1em}{.}
 \label{eq:matrix_dif_fX}
\end{alignat}
From Eq.~\eqref{eq:fX}, we have
\begin{alignat}{1}
 \InsertPDF{matrix_dif_fX2.pdf} ~\raisebox{.1em}{.}
 \label{eq:matrix_dif_fX2}
\end{alignat}

\section{Basic formulas} \label{subsec:basic_rules}

We review some basic formulas that we shall frequently use later.

\subsection{Derivatives of $A$ and $X$}

For any matrix $A$ that is independent of $X$, $\pp{X} A = \zero$,
i.e.,
\begin{alignat}{1}
 \InsertPDF{matrix_dif_dA.pdf}
 \label{eq:matrix_dif_dA}
\end{alignat}
holds.
In what follows, we assume that matrices $A,B,\dots$ are independent of $X$,
unless otherwise mentioned.
Also, from $\pp{X_{i,j}} X_{k,l} = \delta_{i,k} \delta_{j,l}$
(where $\delta_{i,k}$ is the Kronecker delta), we have
$\pp{X} X = \ket{\cup_m} \bra{\cap_n}$, i.e.,
\begin{alignat}{1}
 \InsertPDF{matrix_dif_dX.pdf} ~\raisebox{.1em}{.}
 \label{eq:matrix_dif_dX}
\end{alignat}

\subsection{Rules for sums and products}

The following sum rule holds:
\begin{alignat}{1}
 \pp{X} [f(X) + g(X)] &= \pp{X} f(X) + \pp{X} g(X),
\end{alignat}
which is diagrammatically represented as
\begin{alignat}{1}
 \InsertPDF{matrix_dif_sum.pdf} ~\raisebox{.1em}{.}
 \label{eq:matrix_dif_sum}
\end{alignat}
As for matrix multiplication and tensor products, we have
\begin{alignat}{1}
 \pp{X} f(X) g(X) &= \left[ \pp{X} f(X) \right] g(X)
 + f(X) \left[ \pp{X} g(X) \right], \nonumber \\
 \pp{X} f(X) \ot h(X) &= \left[ \pp{X} f(X) \right] \ot h(X)
 + f(X) \ot \left[ \pp{X} h(X) \right],
\end{alignat}
which are depicted as
\begin{alignat}{1}
 \InsertPDF{matrix_dif_circ.pdf}
 \label{eq:matrix_dif_circ}
\end{alignat}
and
\begin{alignat}{1}
 \InsertPDF{matrix_dif_ot.pdf} ~\raisebox{.1em}{.}
 \nonumber \\
 \label{eq:matrix_dif_ot}
\end{alignat}
Note that we assume that the order of wires does not matter in a diagram.

\subsection{Chain rules}

Given a matrix $X \in \Real^{m \times n}$,
a map $Y:\Real^{m \times n} \to \Real^{m' \times n'}$,
and a map $f:\Real^{m' \times n'} \to \Real^{k \times l}$,
the derivative of $f[Y(X)]$ with respect to $X_{i,j}$ satisfies
\begin{alignat}{1}
 \pp{X_{i,j}} f[Y(X)] &= \sum_{i'=1}^k \sum_{j'=1}^l
 \frac{\partial f[Y(X)]}{\partial Y_{i',j'}}
 \frac{\partial Y_{i',j'}}{\partial X_{i,j}},
\end{alignat}
where $Y_{i',j'} \coloneqq \braket{i'|Y(X)|j'}$.
Thus, $\pp{X} f[Y(X)]$ can be diagrammatically represented by
\begin{alignat}{1}
 \InsertPDF{matrix_dif_chain.pdf} ~\raisebox{.1em}{.}
 \label{eq:matrix_dif_chain}
\end{alignat}

All the formulas presented in this paper can be obtained using the above-mentioned equations.
It is noteworthy that this paper is focused on the matrix differentiation,
but our notation can be easily extended to the case of high-order tensors.

\section{Other basic formulas}

We derive several basic formulas.

\subsection{Derivatives of matrix multiplication and tensor products}

We immediately obtain
\begin{alignat}{1}
 \stackinset{c}{0.05cm}{c}{0.9cm}{\footnotesize\eqref{eq:matrix_dif_circ}}{%
 \stackinset{c}{0.05cm}{c}{0.6cm}{\footnotesize\eqref{eq:matrix_dif_ot}}{%
 \stackinset{c}{0.05cm}{c}{0.3cm}{\footnotesize\eqref{eq:matrix_dif_dA}}{%
 \InsertPDF{matrix_dif_AfXB.pdf}}}} ~\raisebox{.1em}{.}
 \label{eq:matrix_dif_circ_ot}
\end{alignat}

\subsection{Derivative of $X^\T$}

Since $X^\T$ is represented by
\begin{alignat}{1}
 \InsertPDF{matrix_trans.pdf} ~\raisebox{.1em}{,}
 \label{eq:matrix_trans}
\end{alignat}
we have
\begin{alignat}{1}
 \stackinset{c}{-1.38cm}{c}{1.45cm}{\footnotesize\eqref{eq:matrix_trans}}{%
 \stackinset{c}{-1.38cm}{c}{1.15cm}{\footnotesize\eqref{eq:matrix_dif_circ_ot}}{%
 \stackinset{c}{-1.38cm}{c}{-0.69cm}{\footnotesize\eqref{eq:matrix_dif_dX}}{%
 \stackinset{c}{2.05cm}{c}{-0.69cm}{\footnotesize\eqref{eq:matrix_cup_cap_id}}{%
 \InsertPDF{matrix_dif_dXT.pdf}}}}} \nonumber \\
 ~\raisebox{.1em}{.}
 \label{eq:matrix_dif_dXT}
\end{alignat}

\subsection{Derivatives of Hadamard products}

The Hadamard product of $A \in \Real^{m \times n}$ and $B \in \Real^{m \times n}$,
denoted by $A \c B$, is the component-wise product, i.e.,
\begin{alignat}{1}
 A \c B &\coloneqq \sum_{i=1}^m \sum_{j=1}^n \braket{i|A|j} \braket{i|B|j} \ket{i} \bra{j},
\end{alignat}
which is diagrammatically depicted as
\begin{alignat}{1}
 \InsertPDF{matrix_Hadamard.pdf} ~\raisebox{.1em}{.}
 \label{eq:matrix_Hadamard}
\end{alignat}
From Eq.~\eqref{eq:matrix_dif_dXT}, we can readily verify
%
%
\begin{alignat}{1}
 \InsertPDF{matrix_dif_Hadamard.pdf} ~\raisebox{.1em}{.} \nonumber \\
 \label{eq:matrix_dif_Hadamard}
\end{alignat}

\section{Examples}

We will give some concrete examples that are directly derived from
the above basic formulas.

\subsection{Derivatives with respect to column vectors}

\subsubsection{$\displaystyle \pp{\ket{x}} \braket{a|x} = \ket{a}$}
~ \\

Substituting $n = 1$ into Eq.~\eqref{eq:matrix_dif_dX} gives
\begin{alignat}{1}
 \InsertPDF{matrix_dif_d_x.pdf} ~\raisebox{.1em}{.}
 \label{eq:matrix_dif_d_x}
\end{alignat}
Thus, we have
\begin{alignat}{1}
 \stackinset{c}{-0.8cm}{c}{0.3cm}{\footnotesize\eqref{eq:matrix_dif_d_x}}{%
 \InsertPDF{matrix_dif_a_x.pdf}} ~\raisebox{.1em}{.}
 \label{eq:matrix_dif_a_x}
\end{alignat}
Note that $\bra{a}^\T = \ket{a}$ holds since $\ket{a}$ is a real column vector.

\subsubsection{$\displaystyle \pp{\ket{x}} \braket{x|A|x} = (A + A^\T) \ket{x}$}
~ \\

Substituting $n = 1$ into Eq.~\eqref{eq:matrix_dif_dXT} gives
\begin{alignat}{1}
 \InsertPDF{matrix_dif_d_xT.pdf} ~\raisebox{.1em}{,}
 \label{eq:matrix_dif_d_xT}
\end{alignat}
and thus
\begin{alignat}{1}
 \stackinset{c}{-0.68cm}{c}{1.55cm}{\footnotesize\eqref{eq:matrix_dif_d_xT}}{%
 \stackinset{c}{-0.68cm}{c}{1.25cm}{\footnotesize\eqref{eq:matrix_dif_d_x}}{%
 \stackinset{c}{-0.68cm}{c}{-1.0cm}{\footnotesize\eqref{eq:matrix_cup_cap_X}}{%
 \InsertPDF{matrix_dif_x_A_x.pdf}}}}
 \label{eq:matrix_dif_x_A_x}
\end{alignat}
holds.

\subsubsection{Other important examples}
~ \\

We can easily obtain the following formulas
(the proofs are left to the readers)
\footnote{The second line follows from substituting
$u \coloneqq \|\ket{x} - \ket{b}\|_2^2$ into
\begin{alignat}{1}
 \pp{\ket{x}} \sqrt{u} &=
 \frac{\partial u}{\partial \ket{x}} \frac{\partial \sqrt{u}}{\partial u}
 = \frac{\partial u}{\partial \ket{x}} \cdot \frac{1}{2 \sqrt{u}},
\end{alignat}
which is immediately obtained by the chain rule.}:
\begin{alignat}{1}
 \pp{\ket{x}} \|A \ket{x} - \ket{b} \|_2^2 &= 2 A^\T (A \ket{x} - \ket{b}), \nonumber \\
 \pp{\ket{x}} \|\ket{x} - \ket{b}\|_2
 &= \frac{\ket{x} - \ket{b}}{\|\ket{x} - \ket{b}\|_2}.
\end{alignat}

\subsection{Derivatives with respect to matrices}

\subsubsection{$\displaystyle \pp{X} \braket{a|X|b} = \ket{a} \bra{b}$}

\begin{alignat}{1}
 \stackinset{c}{-0.8cm}{c}{0.3cm}{\footnotesize\eqref{eq:matrix_dif_dX}}{%
 \InsertPDF{matrix_dif_aXb.pdf}} ~\raisebox{.1em}{.}
 \label{eq:matrix_dif_aXb}
\end{alignat}

\subsubsection{$\displaystyle \pp{X} \Tr(AX) = A^\T$}

\begin{alignat}{1}
 \stackinset{c}{-0.79cm}{c}{0.3cm}{\footnotesize\eqref{eq:matrix_dif_dX}}{%
 \stackinset{c}{1.85cm}{c}{0.3cm}{\footnotesize\eqref{eq:matrix_cup_cap_X}}{%
 \InsertPDF{matrix_dif_Tr_AX.pdf}}} ~\raisebox{.1em}{.}
 \label{eq:matrix_dif_Tr_AX}
\end{alignat}

\subsubsection{$\displaystyle \pp{X} \Tr(XX^\T) = 2X$}

\begin{alignat}{1}
 \stackinset{c}{-1.11cm}{c}{1.4cm}{\footnotesize\eqref{eq:matrix_dif_dX}}{%
 \stackinset{c}{-1.11cm}{c}{1.1cm}{\footnotesize\eqref{eq:matrix_dif_dXT}}{%
 \stackinset{c}{-1.11cm}{c}{-1.2cm}{\footnotesize\eqref{eq:matrix_cup_cap_X}}{%
 \InsertPDF{matrix_dif_Tr_XXT.pdf}}}} ~\raisebox{.1em}{.}
 \label{eq:matrix_dif_Tr_XXT}
\end{alignat}

\subsubsection{$\displaystyle \pp{X} \Tr(AXBX) = A^\T X^\T B^\T + B^\T X^\T A^\T$}

\begin{alignat}{1}
 \stackinset{c}{-1.05cm}{c}{2.05cm}{\footnotesize\eqref{eq:matrix_dif_dX}}{%
 \stackinset{c}{-1.05cm}{c}{-2.15cm}{\footnotesize\eqref{eq:matrix_cup_cap_X}}{%
 \InsertPDF{matrix_dif_Tr_AXBX.pdf}}} ~\raisebox{.1em}{.}
 \label{eq:matrix_dif_Tr_AXBX}
\end{alignat}

\subsubsection{$\displaystyle \pp{X} X^{-1} = - (\ident \ot X^{-1}) \ket{\cup} \bra{\cap} (\ident \ot X^{-1})$}
~ \\

Letting $Z \coloneqq \pp{X} X^{-1}$ and differentiating $X^{-1} = X^{-1} X X^{-1}$
with respect to $X$ gives
$Z = Z + X^{-1} \frac{\partial X}{\partial X} X^{-1} + Z$.
Thus, we have
\begin{alignat}{1}
 \stackinset{c}{1.32cm}{c}{0.3cm}{\footnotesize\eqref{eq:matrix_dif_dX}}{%
 \InsertPDF{matrix_dif_dXinv.pdf}} \nonumber \\
 ~\raisebox{.1em}{.}
 \label{eq:matrix_dif_dXinv}
\end{alignat}

\subsubsection{$\displaystyle \pp{X} \Tr[(X + A)^{-1}] = - [(X + A)^{-2}]^\T$}
~ \\

Letting $Y \coloneqq X + A$ and using the chain rule, we obtain
\begin{alignat}{1}
 \stackinset{c}{-0.86cm}{c}{3.05cm}{\footnotesize\eqref{eq:matrix_dif_chain}}{%
 \stackinset{c}{-0.86cm}{c}{0.77cm}{\footnotesize\eqref{eq:matrix_dif_dX}}{%
 \stackinset{c}{-0.86cm}{c}{0.47cm}{\footnotesize\eqref{eq:matrix_dif_dXinv}}{%
 \stackinset{c}{-0.86cm}{c}{-2.3cm}{\footnotesize\eqref{eq:matrix_cup_cap_X}}{%
 \InsertPDF{matrix_dif_Tr_XAinv.pdf}}}}} ~\raisebox{.1em}{.}
 \label{eq:matrix_dif_Tr_XAinv}
\end{alignat}

\subsubsection{$\displaystyle \pp{X} \Tr(A \c X) = A \c \ident$}

\begin{alignat}{1}
 \stackinset{c}{0.15cm}{c}{1.65cm}{\footnotesize\eqref{eq:matrix_dif_Hadamard}}{%
 \stackinset{c}{0.15cm}{c}{1.35cm}{\footnotesize\eqref{eq:matrix_dif_dX}}{%
 \InsertPDF{matrix_dif_Tr_H_AX.pdf}}} ~\raisebox{.1em}{.}
 \label{eq:matrix_dif_Tr_H_AX}
\end{alignat}

\subsubsection{$\displaystyle \ppp{\ket{x}}{\bra{x}}(\braket{x|A|x} + \braket{b|x}) = A + A^\T$}

\begin{alignat}{1}
 \stackinset{c}{-2.23cm}{c}{-0.2cm}{\footnotesize\eqref{eq:matrix_dif_x_A_x}}{%
 \stackinset{c}{-2.23cm}{c}{-0.5cm}{\footnotesize\eqref{eq:matrix_dif_a_x}}{%
 \stackinset{c}{-2.23cm}{c}{-2.3cm}{\footnotesize\eqref{eq:matrix_dif_d_x}}{%
 \stackinset{c}{-2.23cm}{c}{-2.6cm}{\footnotesize\eqref{eq:matrix_cup_cap_id}}{%
 \InsertPDF{matrix_dif_Hessian.pdf}}}}} ~\raisebox{.1em}{.}
 \label{eq:matrix_dif_Hessian}
\end{alignat}
This formula shows that the Hessian matrix of the quadratic function
$\braket{x|A|x} + \braket{b|x} + c$ with
$A \in \Real^{m \times m}$, $\ket{b} \in \Real^m$, and $c \in \Real$
is $A + A^\T$.

\subsubsection{Other important examples}
~ \\

We can easily obtain the following formulas (the proofs are left to the readers):
\begin{alignat}{1}
 \pp{X} \Tr(AXB) &= A^\T B^\T, \nonumber \\
 \pp{X} \Tr(X \ot X) &= (2 \Tr X) \ident, \nonumber \\
 \pp{X} \braket{a|X^\T CX|b} &= CX\ket{b}\bra{a} + C^\T X\ket{a}\bra{b}, \nonumber \\
 \pp{X} \Tr(X^k) &= k(X^{k-1})^\T, \nonumber \\
 \pp{X} \Tr(AX^k) &= \sum_{s=0}^{k-1} (X^s A X^{k-1-s})^\T, \nonumber \\
 \pp{X} \Tr(AX^{-1}B) &= -(X^{-1}BAX^{-1})^\T,
\end{alignat}
where $k$ is a natural number.

\section{Conclusion}

We introduced a diagrammatic notation for matrix differentiation.
We demonstrated through some interesting examples that
our notation makes it possible to easily and intuitively calculate matrix differentiation.

\section*{Acknowledgment}

I am grateful to O. Hirota for support.

\bibliographystyle{myieeetr}


\end{document}